\documentstyle[galley,epsf]{mn}
\newif\ifAMStwofonts
\ifoldfss
  \ifCUPmtlplainloaded \else
    \NewTextAlphabet{textbfit} {cmbxti10} {}
    \NewTextAlphabet{textbfss} {cmssbx10} {}
    \NewMathAlphabet{mathbfit} {cmbxti10} {} % for math mode
    \NewMathAlphabet{mathbfss} {cmssbx10} {} %  "   "    "
  \fi
  \ifAMStwofonts
    \ifCUPmtlplainloaded \else
      \NewSymbolFont{upmath} {eurm10}
      \NewSymbolFont{AMSa} {msam10}
      \NewMathSymbol{\upi}     {0}{upmath}{19}
      \NewMathSymbol{\umu}     {0}{upmath}{16}
      \NewMathSymbol{\upartial}{0}{upmath}{40}
      \NewMathSymbol{\leqslant}{3}{AMSa}{36}
      \NewMathSymbol{\geqslant}{3}{AMSa}{3E}

    \fi
  \fi
\fi % End of OFSS

\ifnfssone
  \newmathalphabet{\mathit}
  \addtoversion{normal}{\mathit}{cmr}{m}{it}
  \addtoversion{bold}{\mathit}{cmr}{bx}{it}
  \newmathalphabet{\mathbfit} % math mode version of \textbfit{..}
  \addtoversion{normal}{\mathbfit}{cmr}{bx}{it}
  \addtoversion{bold}{\mathbfit}{cmr}{bx}{it}
  \newmathalphabet{\mathbfss} % math mode version of \textbfss{..}
  \addtoversion{normal}{\mathbfss}{cmss}{bx}{n}
  \addtoversion{bold}{\mathbfss}{cmss}{bx}{n}
  \ifAMStwofonts
    \ifCUPmtlplainloaded \else
      \UseAMStwoboldmath
      \makeatletter
      \new@mathgroup\upmath@group
      \define@mathgroup\mv@normal\upmath@group{eur}{m}{n}
      \define@mathgroup\mv@bold\upmath@group{eur}{b}{n}
      \edef\UPM{\hexnumber\upmath@group}
      \new@mathgroup\amsa@group
      \define@mathgroup\mv@normal\amsa@group{msa}{m}{n}
      \define@mathgroup\mv@bold\amsa@group{msa}{m}{n}
      \edef\AMSa{\hexnumber\amsa@group}
      \makeatother
      \mathchardef\upi="0\UPM19
      \mathchardef\umu="0\UPM16
      \mathchardef\upartial="0\UPM40
      \mathchardef\leqslant="3\AMSa36
      \mathchardef\geqslant="3\AMSa3E
    \fi
  \fi
\fi % End of NFSS release 1

\ifnfsstwo
  \DeclareMathAlphabet{\mathbfit}{OT1}{cmr}{bx}{it}
  \SetMathAlphabet\mathbfit{bold}{OT1}{cmr}{bx}{it}
  \DeclareMathAlphabet{\mathbfss}{OT1}{cmss}{bx}{n}
  \SetMathAlphabet\mathbfss{bold}{OT1}{cmss}{bx}{n}
  \ifAMStwofonts
    \ifCUPmtlplainloaded \else
      \DeclareSymbolFont{UPM}{U}{eur}{m}{n}
      \SetSymbolFont{UPM}{bold}{U}{eur}{b}{n}
      \DeclareSymbolFont{AMSa}{U}{msa}{m}{n}
      \DeclareMathSymbol{\upi}{0}{UPM}{"19}
      \DeclareMathSymbol{\umu}{0}{UPM}{"16}
      \DeclareMathSymbol{\upartial}{0}{UPM}{"40}
      \DeclareMathSymbol{\leqslant}{3}{AMSa}{"36}
      \DeclareMathSymbol{\geqslant}{3}{AMSa}{"3E}
    \fi
  \fi
\fi % End of NFSS release 2

\ifCUPmtlplainloaded \else
  \ifAMStwofonts \else % If no AMS fonts
    \def\upi{\pi}
    \def\umu{\mu}
    \def\upartial{\partial}
  \fi
\fi

\begin{document}
\title[R CrB 2003 minimum]
   {R\,Coronae Borealis at the 2003 Light Minimum\thanks{Based on
observations obtained with the Hobby-Eberly Telescope, which is a joint
project of the University of Texas at Austin, the Pennsylvania State
University, Stanford University, Ludwig-Maximilians-Universit\"{a}t
M\"{u}nchen, and Georg-August-Universit\"{a}t G\"{o}ttingen.}}

\author[N. Kameswara Rao, David L. Lambert, \& Matthew D. Shetrone]
       {N. Kameswara Rao$^1$,  David L. Lambert$^2$, \& Matthew D. Shetrone$^3$\\
       $^1$Indian Institute of Astrophysics, Bangalore 560034, India\\
       $^2$The W.J. McDonald Observatory, University of Texas, Austin, TX 78712-1083, USA\\
$^3$The W.J. McDonald Observatory, Fort Davis, TX 79734-9875}
\date{Accepted 
      Received ; 
      in original form  }
                                                                                  
\pagerange{\pageref{firstpage}--\pageref{lastpage}}
\pubyear{}

\maketitle

\label{firstpage}

\begin{abstract}

A set of five high-resolution optical spectra of R CrB obtained in 2003 March
is discussed. At the time of the first spectrum (March 8) the star
was at V = 12.6, a decline of more than six magnitudes. By March 31, the date of the
last observation, the star at V = 9.3 was on the recovery to maximum
light (V = 6). The 2003 spectra 
are compared with the extensive
collection of spectra from the 1995-1996 minimum presented  previously.
Spectroscopic features common to the two minima include the familiar
ones also seen in spectra of other RCBs in decline: sharp emission
lines of neutral and singly-ionized atoms, broad emission lines including
He\,{\sc i}, [N\,{\sc ii}] 6583 \AA,  Na D, and Ca\,{\sc ii} H \& K lines,
and
blueshifted aborption
lines of Na D, and K\,{\sc i} resonance lines. Prominent differences between the
2003 and 1995-1996 spectra are seen. The broad Na D and Ca H \& K lines in 2003
and 1995-1996 are centred approximately on the mean stellar velocity. The 
2003 profiles
are fit by a single Gaussian but in 1995-1996 two Gaussians separated by about
200 km s$^{-1}$ were required. However, the He\,{\sc i} broad emission lines are
fit by a single Gaussian   at all times;  the emitting
He and Na-Ca atoms are probably not colocated.  The C$_2$ Phillips 2-0 lines are
detected as sharp absorption  lines and the C$_2$ Swan band lines as sharp emission
lines in 2003 but in 1995-1996 the Swan band emission lines were broad and the
Phillips lines were undetected. The 2003 spectra show C\,{\sc i} sharp
emission lines at minimum light with a velocity changing in five days by
about 20 km s$^{-1}$ when the velocity of `metal' sharp lines is unchanged; the
C\,{\sc i} emission may arise from shock-heated gas. Reexamination of spectra
obtained at maximum light in 1995 shows extended blue wings to strong lines
with the extension dependent on a line's lower excitation potential; this is
the signature of a stellar wind, also revealed by published observations of the
He\,{\sc i} 10830 \AA\ line at maximum light. Changes in the cores of the
resonance lines of Al\,{\sc i} and Na D (variable blue shifts) and
the Ca\,{\sc ii} IR lines (variable blue and red shifts) suggest complex
flow patterns near the photosphere. The spectroscopic differences at the two mimima show the
importance of continued scrutiny of the declines of R CrB (and
other RCBs). Thorough understanding of the outer atmosphere and circumstellar
regions of R CrB will require such continued scrutiny.

\end{abstract}


\section{Introduction}

R Coronae Borealis stars (here, RCBs), a class with about 30 Galactic members,
are H-poor  supergiants that decline
in brightness unpredictably and rapidly  by up to  8 magnitudes
to remain at or near minimum light for several weeks to
months.
Although it is generally accepted that the declines
are due to formation of an obscuring cloud of carbon soot
(Loreta 1934; O'Keefe 1939),
  many questions remain unanswered about the
formation and evolution of the dust  clouds (Clayton 1996).
 The discovery
that RCBs possess an
infrared excess even at maximum light suggested that a swarm of dust clouds
is a permanent feature (Feast 1975, 1979, 1986).
Dust forms over a part of the star. If it forms over the Earth-facing
hemisphere, a decline ensues. Often, the dust  forms at other
locations and then no (or a weak) decline will be seen.
Recently, dust clouds have been directly
imaged in the infrared at 2.17 and 4.05 microns around the RCB star
 RY Sgr (de Laverny \& M\'{e}karnia 2004).

Spectroscopic observations of RCBs in decline, when the stellar
photosphere is heavily obscured, have the potential to
provide novel and unique
 information about the outer stellar atmosphere and circumstellar
region. To date,
 high resolution spectroscopic optical observations of RCBs
 in deep declines
have been  limited
in number and scope.
We are attempting to gather
spectra of RCBs in deep declines, as opportunities arise.
Our goals are twofold: (i) to obtain at least one high-resolution optical
spectrum of as many RCBs in decline as possible, and (ii) to observe several
deep declines of the brightest RCBs with an emphasis on the prototype
R CrB, which is readily accessible from the McDonald Observatory.

Here, we report on spectra obtained during the 2003 decline of
R CrB and, in particular, we compare and contrast
the spectra with those obtained during our intensive coverage
of the star's 1995-1996 decline (Rao et al. 1999).
The 2003 minimum was
well separated in time from the preceding one that occurred in
2001. (A brief spectroscopic account of the recovery from the 2001 minimum
has been given by Kipper (2001).)
 Our sequence of five
high-resolution spectra of
R CrB was obtained in 2003 March  during and following the deep minimum.

The light curve obtained from the AAVSO and Efimov (2004 -private communication)
shows that the star started to decline on or
about 2003 February 9.
The decline was rapid reaching V $\simeq$ 12.9 by March 1. The star was
 fainter than
 V $\simeq$ 12.5  for about two weeks from late February to March 17 and then
brightened gradually reaching V $\simeq$ 9.4
by March  31.
The visual light curve from onset of the
decline through to recovery almost to maximum light is shown
in Figure 1 with the times at which our five spectra were taken
indicated by arrows.
The first two
spectra were
acquired when the star was at minimum light at V $\simeq$ 12.5.
The third spectrum was taken in the early stages of
recovery to maximum light with the star at V $\simeq 10.7$.  The final
pair of spectra was taken when the star had brightened to  V $\simeq$ 9.3.
At maximum, R CrB is at V = 6. During the 1995-1996 decline, R CrB faded to
V = 13.6.

\begin{figure}
\epsfxsize=8truecm
\epsffile{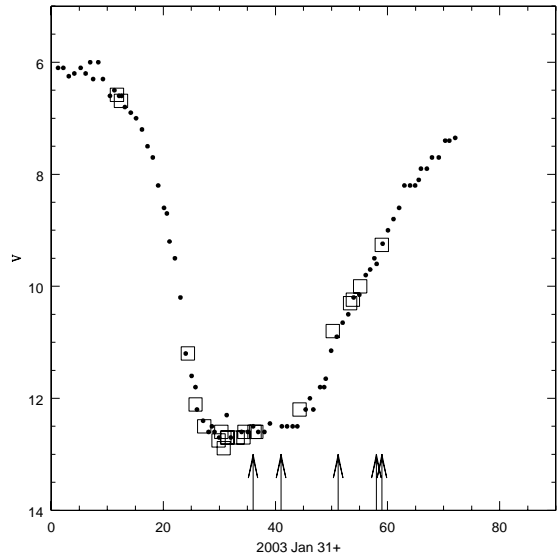}
\caption{The light curve of R CrB for the 2003 minimum.
Magnitude estimates are from the AAVSO (circles) and Efimov (2004, private communication)
(open squares). The x-axis is in days with the origin at 2003 January 31.
Arrows denote
the times of our five spectroscopic observations }
\end{figure}

\section{Observations}

Our first two spectra were
taken with the queue-scheduled 9.2 meter Hobby-Eberly Telescope (HET) and
its High-Resolution Spectrometer (Tull 1998) 
These HET spectra
are at a resolving power of $R = \lambda/d\lambda \simeq
60000$ and provide full wavelength coverage from
5320 \AA\ to 7330 \AA.
                                                                                  
\begin{table*}
\centering
\begin{minipage}{90mm}
\caption{\Large Observations of R CrB in 2003}
\begin{tabular}{llrr}
\hline
                                                                                  
Date & Julian Date & Magnitude & Telescope\\
(UT) & (2452000+) \\
\hline
March 8 & 706.794 & 12.6 & HET\ 9.2m\\
March 13 & 711.794 & 12.5 & HET\ 9.2m\\
March 23 & 721.780 & 10.7 & McDonald\ 2.7m\\
March 30 & 728.857 & 9.4 & McDonald\ 2.7m\\
March 31 & 729.864 & 9.3 & McDonald\ 2.7m\\
\hline
\end{tabular}
\label{default}
\end{minipage}
\end{table*}

 The remaining three spectra
were acquired with the McDonald Observatory's  Harlan J. Smith 2.7 meter telescope and
the {\it 2dcoud\'{e}} cross-dispersed echelle spectrometer (Tull et al. 1995).
These  observations, which were obtained when R CrB was
recovering from the
minimum,
cover the interval 3700 \AA\ to 10000 \AA\ with complete spectral
coverage shortward of about 5500 \AA.
The resolving power was  $R
\simeq 60,000$, as measured from the thorium lines in the
Th-Ar hollow cathode comparison spectrum.

\section{Spectroscopic signatures of a RCB in decline}

For a RCB in decline, emission
lines dominate the optical spectrum.
Two broad classes of emission lines are present:
 a rich set of sharp lines (FWHM $\sim 12$
km s$^{-1}$), and a sparse and diverse set of broad lines (FWHM $\sim 300$ km s$^{-1}$)
(Herbig 1949; Payne-Gaposchkin 1963; Alexander et al. 1972).
Sharp lines are primarily  low-excitation transitions
of singly-ionized  and neutral metals that
appear very early in the decline and only disappear late in the
recovery to maximum light (Rao et al. 1999).
 The broad lines, which are seen only when a RCB has
faded by several magnitudes, may include lines of the He\,{\sc i} triplet series,Ca\,{\sc ii} H and K, K\,{\sc i} resonance lines at 7664\AA\ and 7699 \AA,
Na\,{\sc i} D lines, [O\,{\sc ii}],and [N\,{\sc ii}] lines, i.e., a mix of high and
low excitation lines with  differing broad profiles (Rao et al. 1999) suggesting
that there may be perhaps three regions responsible broad
emission lines.
                                                                                 
The photospheric absorption line spectrum also changes during a decline.
 In deep minima, the photospheric absorption
lines are `veiled', i.e., the lines become
very shallow and broad.
New absorption features may also appear.
 Broad blueshifted absorption components
have been seen to accompany commonly  the Na  D lines, and  occasionally
the  K\,{\sc i} 7664 \AA\ and 7699 \AA\ resonance lines,
and the Ca\,{\sc ii} H and K lines.\footnote{Whitney,
  Dupree \& Zucker (1993) claim absorption components for
the Ca\,{\sc ii} K line at -240 and -180 km s$^{-1}$ even at maximum light.
These features correspond to the position of Fe\,{\sc i} lines at 3930.3 \AA\
and 3931.1 \AA. Moreover,
these blueshifted absorption features  are
not obvious in H line.}
The Na D absorption components appear especially at and following minimum
light.
% Blue-shifted absorption accompanies the He\,{\sc i} 10830 \AA\
%line  at
%maximum light (Clayton, Geballe, \&
%Bianchi  2003) but it is by no means clear that this absorption has anything to do with
%that seen at
% minimum is light.

Descriptions of the emission and absorption lines in
decline rest
largely on high resolution spectroscopic data gathered at
 declines of the two
brightest RCBs  -- R CrB and RY Sgr (Payne-Gaposchkin 1963;
Alexander et al. 1972; Cottrell, Lawson, \& Buchhorn 1990;
 Rao et al. 1999; Kipper 2001). Other RCBs
observed on  one or a few occasions  provide useful information
toward establishing a model of a RCB's outer atmosphere and circumstellar
envelope. Published reports of spectra of other RCBs in decline refer
to S Aps (Goswami et al. 1997),
V854 Cen
(Whitney et al. 1992 ;
 Rao \& Lambert 1993, 2000; Skuljan \& Cottrell 2002), and
 UW Cen (Rao, Reddy, \& Lambert 2004).

The ubiquity of the  sharp and broad emission and the blue-shifted absorption Na D lines
across the sample of RCBs is unknown at present. Probably,
the sharp emission lines of ionized and neutral metals are a common
feature of all declines of all RCBs (Skuljan \& Cottrell 2004).\footnote{Emission
lines of
C\,{\sc i} and O\,{\sc i} first seen early in R CrB's 1995 decline
may have been missed in other declines of R CrB and other RCBs simply
for lack of appropriate observations at early times.}
Since few  RCBs have  been observed in deep minima,  reported
sightings of  broad lines are rare.
 One observation worthy of particular
note, if applicable to all RCBs, is Whitney et al.'s discovery that the Na D broad
emission from V854 Cen is unpolarized at a time when the continuum is
markedly polarized. This strongly suggests that the Na D broad emission is
not viewed through the dusty cloud responsible for the decline.

Detailed descriptions of  spectra of the same RCB at different declines
are rare in the literature.
Here, we compare the 2003 spectra with those from our extensive

\section{Emission Lines}

\subsection{The Broad Emission Lines}

Broad emission lines seen in 2003 include
lines of He\,{\sc i} at  3889 \AA, 5876 \AA, and  7065 \AA,
the Na\,{\sc i} D lines, the Ca\,{\sc ii} H \& K lines,
 and the [N\,{\sc ii}]  line at 6583 \AA.
These are the strongest lines anticipated from our earlier
work on R CrB (Rao et al. 1999) and V854 Cen (Rao \& Lambert 1993).
Other lines would surely have been seen had the spectra been of higher
S/N ratio.

\subsubsection{He\,{\sc i} lines}

\begin{figure}
\epsfxsize=8truecm
\epsffile{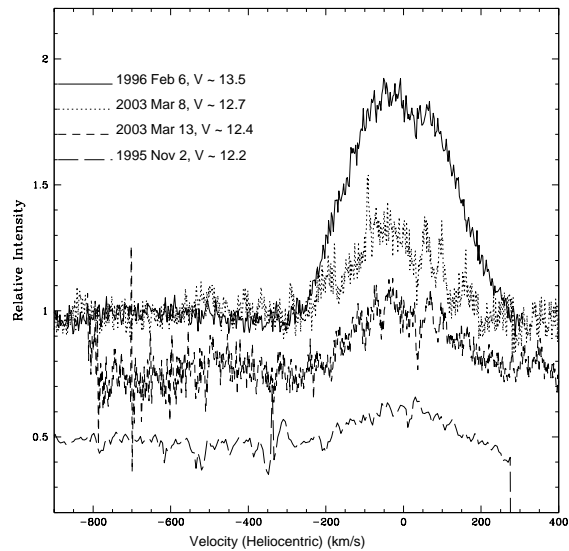}
\caption{The He\,{\sc i} line at  7065 \AA\  on 2003 March 8 and 13 flanked
by the profiles observed on 1995 November 2 and 1996 February 6.
The spectra are ordered top to bottom by the V magnitude of the star at the
time of observation. The V magnitudes are given on the figure. In this and
subsequent figures  a constant  shift in relative intensity has been 
applied to the spectra for clarity.}
\end{figure}

Broad emission due to He\,{\sc i} at 7065 \AA\ is clearly seen in the
HET spectra for March 8 and 13.
The 5876 \AA\ line is weakly
present in the March 8 spectrum.
 Figure 2 compares the 7065 \AA\ profiles with two
observations made during minimum light in the 1995-1996
decline with all spectra normalized to a local continuum.
 The He\,{\sc i} line  appears strongest (relative to the local continuum)
 in  the 1996 February 6
observation probably 
because the star was then a magnitude 
fainter (V $\simeq 13.5$) than in 2003 March;
% Rao et al. (1999) suggested that the line flux
%was approximately constant.
The 7065 \AA\ line is not seen in the spectra from March 23 and 31. This
is surely because the star had brightened by two magnitudes or more so
reducing the line to continuum contrast.

At the minimum of the 1995-1996 decline, the 7065 \AA\ line appears 
with a quasi- parabolic profile
(see the 1996 February 6 profile in Figure 2).
The apparent irregularities in the 2003 profiles
are attributable to the lower S/N ratio of these
spectra which  may be contaminated with telluric H$_2$O lines. The
1995-1996 spectra were ratioed with spectra of a hot star to
remove the contaminating H$_2$O lines.
 The radial velocity of the 2003 emission peak (approximately the
 centre of the profile) is $-$38 km s$^{-1}$ (Table 2). The emission extends
 from $-$250 km s$^{-1}$
to $+$190 km s$^{-1}$ with a width (FWHM) of 251 km s$^{-1}$.  In 1996, the peak
emission was at $-13$ km s$^{-1}$ and the FWHM of the parabolic-like profile
was 271 km s$^{-1}$.

%\begin{figure}
%\epsfxsize=8truecm
%\epsffile{plot3rcb03feb17b.ps}
%\caption{The He\,{\sc i} line at  3889 \AA\  on 2003 March 23 (solid line)
%and on 1995 May 9 (dotted line) from the 1995-1996
%minimum. Note the three prominent sharp lines -- the Ti\,{\sc ii} lines at
%3900 and 3913 \AA\ and the Si\,{\sc i} line at 3905 \AA. Many weaker
%sharper lines are seen in both spectra.}
%\end{figure}

\begin{table*}
\centering
\begin{minipage}{90mm}
\caption{\large Radial Velocities (km s$^{-1}$) of various features at the 2003 minimum of R CrB.}
\begin{tabular}{lrrrrr}
\hline
\multicolumn{1}{c}{Feature}&\multicolumn{1}{c}{}&\multicolumn{1}{c}{ }&\multicolumn{1}{c}{Date}&
\multicolumn{1}{c}{} &\multicolumn{1}{c}{}\\
\cline{3-6} \\
&&\multicolumn{1}{c}{ March 8 }& \multicolumn{1}{c}{March 13 }
& \multicolumn{1}{c}{March 23 } & \multicolumn{1}{c}{March 31}\\
\hline
Absorption lines&       &   30.4 (17)$^a$ & 16.9 ( 9) &  17.8 (42)& 13.5 (45)\\
           &            &  $\pm$2.3   & $\pm$1.6  & $\pm$3.4  & $\pm$1.5 \\
Sharp emissions&        &   19.1 (79) & 22.1 (80) &  23.1 (29)& 22.1 (20) \\
           &            &  $\pm$1.5   & $\pm$2.7  & $\pm$1.7  & $\pm$2.7  \\
           &[O\,{\sc i}] &   18.3      & 23.8      &  23.4    &          \\
           &[Ca\,{\sc ii}]&   19.5 ( 2) & 19.3 ( 2) &  23.0 ( 2)& 24.2 ( 2) \\
           &            &             &           &           &            \\
Shell absorption&Na\,{\sc i} D&  $-$93      & $-$96     & $-$115    &$-$132    \\
           &            &             &           &           &            \\
Broad He\,{\sc i} emission &  &             &           &           &            \\
           & 3889\AA\   &             &           & $-$38.6   &            \\
           & 5876\AA\   &  $-$37.7    &           &           &            \\
           & 7065\AA\   &  $-$38.4    & $-$38     &           &            \\
           &            &             &           &           &            \\
C$_2$ (Phillips)&        &             &           &           &            \\
  (2-0) absorption & &             &           &  9.6  ( 6)&            \\
\hline
\end{tabular}
$^a$The values given in parenthesis refer to the number of lines used\\
\end{minipage}
\end{table*}

The He\,{\sc i} 3889 \AA\ line is outside the wavelength range recorded on the HET
spectra but it is prominent in the McDonald spectrum obtained on 2003 March 23
(Figure 3) but
has almost disappeared by March 30-31.
The 3889 \AA\ line appears to be symmetrical apart from the
superposition of the sharp lines.
%Figure 3 shows the
%2003 March 23  spectrum around 3889 \AA\ with a spectrum from the 1995-1996
%when the star was at about the same magnitude (V $\simeq 10.4$).
Figure 3 shows the 2003 March 23 profile with two from the 1996
decline.
The blue wings of the 2003 March 23 and the 1996 May 6  profiles
 are blended with one and possibly
two sharp lines. The bluemost of this pair is the Fe\,{\sc i} line at
3886.3 \AA. The other sharp
line may be a Fe\,{\sc i} line at 3887.1 \AA.
The profiles for 3889 \AA\ and 7065 \AA\ profiles on 1996 February 6 are quite
similar; note especially the similar shape at the lines' peak.
The radial velocity for the 2003 He\,{\sc i} line is estimated to be
$-39$ km s$^{-1}$, a similar value to that from the 7065 \AA\ line.
The radial velocity was -14 km s$^{-1}$ for the 1995-1996 minimum.
                                                                                 
\begin{figure}
\epsfxsize=8truecm
\epsffile{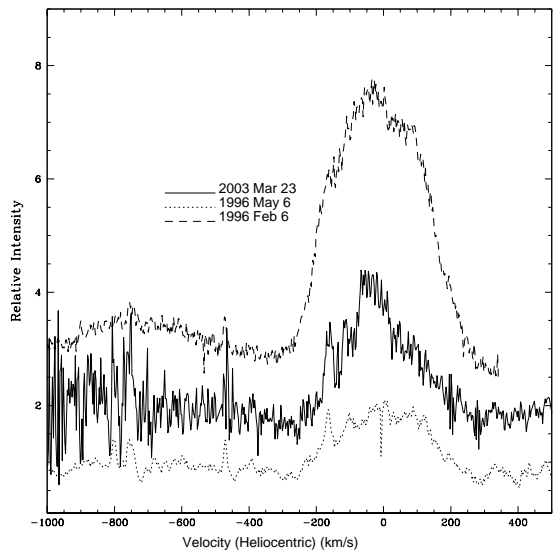}
\caption{The He\,{\sc i} line at  3889 \AA\  on 2003 March 23 (solid line),
1995 May 5 (dotted line) and  on 1996 February 6 (dashed line).}
\end{figure}

%\begin{figure*}
%\epsfxsize=18truecm
%\epsffile{rcrb2003_fig4.eps}
%\caption{The He\,{\sc i} line at  3889 \AA\  on 2003 March 23
%with the profile observed on 1996 Feb 6 during the deepest part of
% the 1995-1996 minimum.}
%\end{figure*}

 The flux in a He\,{\sc i} line may be estimated from the UBVRI magnitudes
 interpolated to the dates of our observation and  the flux
 calibration suggested by Wamsteker (1981). The flux in the  7065 \AA\
line
 is 6.0$\pm$0.1 $\times 10^{-14}$ erg cm$^{-2}$ s$^{-1}$ on March
8 and 13.
The flux in the
 3889 \AA\ line is  4.5$\pm0.6 \times 10^{-13}$ erg cm$^{-2}$ s$^{-1}$
on March 23.
Fluxes similarly estimated for the  1995-1996 minimum were
 7.2$\pm0.2 \times 10^{-14}$ erg cm$^{-2}$ s$^{-1}$ for 7065 \AA\ and
 3.8$\pm0.3 \times 10^{-13}$ erg cm$^{-2}$ s$^{-1}$
for the 3889 \AA\ line. These estimates show that the He\,{\sc i}
fluxes have changed  little between the two minima. Rao et al.
(1999) noted that little change occurred over the time the
lines were visible  during the 1995-1996 minimum.
It would appear that the flux emitted in the He\,{\sc i}
lines may be independent of the ongoing decline and an intrinsic characteristic
of the circumstellar shell.

\subsubsection{Na\,{\sc i}  D lines}

Each of the Na D lines consists of a sharp component blended with
a broad component, a combination seen in the 1995-1996 decline and
for other RCBs caught in decline. The blue wing of D1's broad component is blended
with the red wing of D2's broad component. Figure 4 shows the
Na D profiles for 2003 March 8, 13, 23, and 31 plotted
to  bring out
the growth of the blue-shifted absorption.

\begin{figure}
\epsfxsize=8truecm
\epsffile{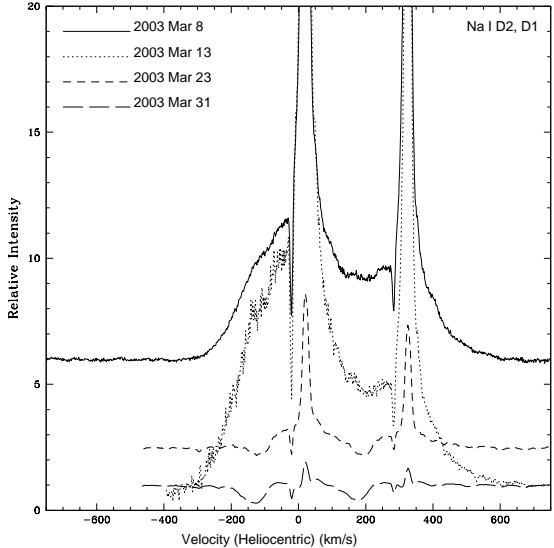}
\caption{Na D profiles in 2003 March showing the sharp and broad
emission components and the development of the high-velocity blue-shifted
absorption component.}
\end{figure}

The blue-shifted absorption is first suspected in the March 8 spectrum
where it is superposed at $-93$ km s$^{-1}$
on the blue wing of the broad D2 emission. In subsequent spectra, this
absorption becomes more  prominent and  shifts
 blueward reaching a velocity of $-132$ km s$^{-1}$ on March 31.
(This absorption component is also seen
 in the K\,{\sc i} 7664 \AA\ and 7699 \AA\
resonance lines.)
The velocities refer to the position of absorption maximum.
As  in the 1995-1996 and other minima, this absorption
appears either soon after minimum light or at the beginning of
the recovery to maximum light. At some minima, blue-shifted
absorption is also seen in the Ca\,{\sc ii} H and K lines,
as shown in Figure 5 where a spectrum from 1998 December 28 shows
clearly the absorption in the H line. The 1998 minimum was shallow,
a drop of only 2.5 magnitudes, but of long duration. Our spectrum was
obtained 126 days following the initial decline when the star
was 1.2 magnitudes below maximum light. This counterpart in
the H line to the Na D absorption was not seen in our limited
set of 2003 spectra.

\begin{figure}
\epsfxsize=8truecm
\epsffile{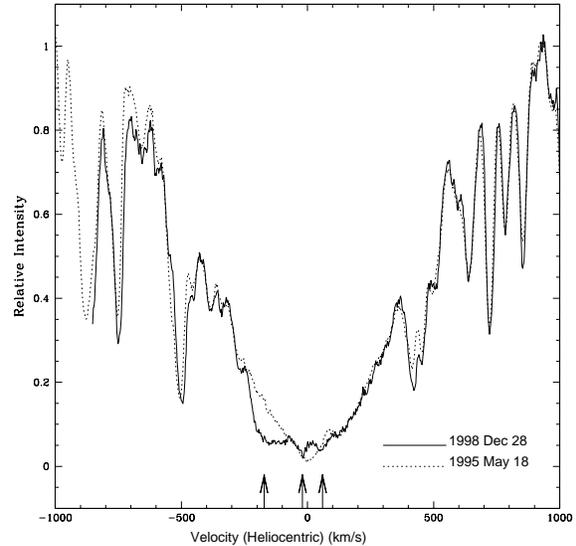}
\caption{The Ca\,{\sc ii} H line in  1995 May 18 at maximum light
and 1998 December 28  during recovery from minimum light. Note the
blue-shifted shell absorption in the H line in the latter
spectrum.}
\end{figure}

There is a distinct difference in the profiles of the broad emission
at the two declines (Figure 6).
The broad emission of D1 and D2 in the 2003 spectra overlaps in the wavelength interval
between the lines. A decomposition of the total emission into
two Gaussians is possible.
 For the 2003 March 8 spectrum, the equivalent
widths are 35.7 \AA\ for D2 and 16.9 \AA\ for D1.
Their ratio is 2.1 which equals to within the measurement
uncertainties the ratio of 2.0 for optically
thin lines.
 The mean (FWHM) width of
the Gausssian fits to D1 and D2 is 250 km s$^{-1}$.
The mean
 radial velocity
is $0\pm7$ km s$^{-1}$.
    (The profiles of 2003  Na \,{\sc i} D broad emission lines are similar to
those at  the 1989 minimum of R CrB (Lambert, Rao \& Giridhar 1990).)

%\begin{figure*}
%\epsfxsize=18truecm
%\epsffile{nkrplot6feb21.ps}
%\caption{Na D profiles in 2003 March 8 compared with the profiles
%observed on 1995 Nov 2 (scaled by the magnitude difference).Note the
%extra emission peak on the redside to the broad
%emission components in 1995 minimum.}
%\end{figure*}

    The profiles of D1 and D2 broad emissions at  
the  1995-1996 minimum
  are quite different from those observed in 2003.
 The blended emission between the
 two lines as well as on the red side of D1 is much more pronounced
 than in 2003 minimum.
While the broad emission in 2003 is well fitted by a single Gaussian, 
 each D line in the 1995-1996
profile requires
 two Gaussians
 separated by 213$\pm5$ km s$^{-1}$ with the red component  stronger than
the blue component by about 50 \%.  The blue component has a larger FWHM
(220 km s$^{-1}$) than the red component (180 km s$^{-1}$). The mean
velocity of the pair of Gaussians is about $-2$ km s$^{-1}$.
This decomposition of the blended Na D lines into a pair of Gaussians
for each line is confirmed by observations
in 1995-1996 of the 
K\,{\sc i} resonance lines at 7664 \AA\ and 7699 \AA\  for which
self-blending is nonexistent.
 In contrast to
the broad emission He\,{\sc i} profiles, the broad emission in the Na D lines is unchanged in
radial velocity but changed in profile between 1995-1996 and 2003.

%A similar red to blue asymmetry was seen in
% the Ca \,{\sc ii} H \& K lines in the 1995-1996
% minimum.

\begin{figure}
\epsfxsize=8truecm
\epsffile{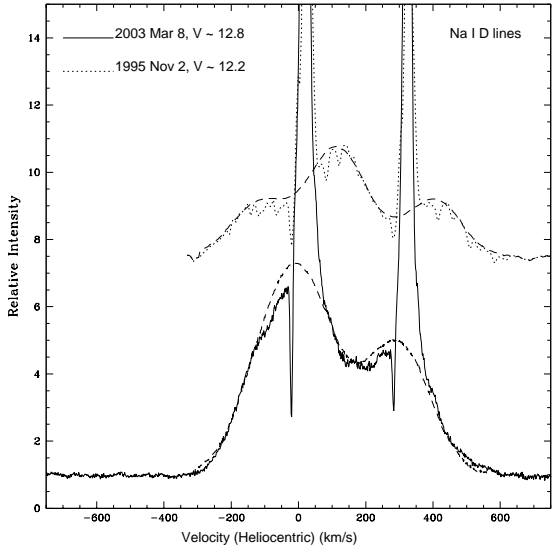}
\caption{The Na D profiles for 2003 March 8 and 1995 November 2 showing the
broad and sharp emission of the D1 and D2 lines. The dashed lines show
Gaussian fits to Na D broad emission  for 2003 March 8 (one component)and
for 1995 November 2 (two components)}
\end{figure}

Although the Na D profile changed, the broad emission's
flux  is similar for the two minima.
  For the first of the 2003 spectra,
 the combined flux of D1 and D2  is 1.7
  $\pm0.2 \times 10^{-12}$ erg cm$^{-2}$ s$^{-1}$ . Estimates for
  1995-1996 minimum are 1.4$\pm0.2 \times 10^{-12}$ erg cm$^{-2}$ s$^{-1}$
  for 1995 November 2 and 1.6$\pm0.2 \times 10^{-12}$ erg cm$^{-2}$ s$^{-1}$
  for 1996 February 6.

%{\bf FWHM of the single Gaussian vs the FWHM of the two 1996 Gaussians?}
% The Gaussian fits are made to the observed profiles with the following
%considerations . Matching of the both red and blue wings of broad lines.
%Even though the D1 and D2 lines are blended the seperation of the
%Gaussian peaks are maintained at the expected value of 6 A. The flux
%ratio of red to blue peak of D2 and D1 are kept at same value. Some
%allowance for overlying absorptions (shell and ISM as well as terrestrial
%water vapour lines) is made.
%      The FWHM of the single Gaussian fits to the profiles on 2003 Mar 8
%are 271 and 220 km s$^{-1}$ for D2 and D1 respectively with a total
%flux for both lines estimated to be 1.66 $\pm0.2 \times 10^{-12}$ erg cm$^{-2}$ s$^{-1}$. The line centres occur at -10.3 and 6.7 km s$^{-1}$.
%Where as in the profiles in 1995 minimum are fitted with two component
%Gaussians . On 1995 Nov 2 spectrum the FWHM for the blue and
%red components of D2 are
%204 and 185 km s$^{-1}$ respectively and the flux ratio of red to blue
%is 1.48, while the FWHM for blue and red components of D1 estimated as
%240 and 181 km s$^{-1}$ respectively with a flux ratio of 1.51.
%The total flux in both lines of both components is estimated as
%1.67 $\pm0.2 \times 10^{-12}$ erg cm$^{-2}$ s$^{-1}$  ver similar to
%the 2003 Mar 8 value. The blue and red peaks occur at -108 and 105
%km s$^{-1}$.

\subsubsection{Ca\,{\sc ii}  lines}

 The Ca$^+$ ion may be expected to  contribute
 the 3968 \AA\ (H) and 3933 \AA\ (K) resonance
lines, the infrared (IR) triplet lines at 8542 \AA, 8662 \AA\ and 8498 \AA,
and the forbidden lines at 7291 \AA\ and 7323 \AA. Unfortunately, our HET
2003 spectra  do
not cover the regions of H and K lines and the  infrared triplet lines.
The forbidden lines are seen at minimum light spectra as sharp emissions
 with an equivalent width ratio of 7291 \AA\ to 7323 \AA of 2 to 1, as anticipated.
A broad component to the forbidden lines with the flux seen at
the one magnitude deeper minimum in 1996 would be masked by the
brighter continuum
of the 2003 minimum.

    The H \& K,  infrared, and forbidden lines were  covered in the McDonald
spectra obtained  in the recovery phase   with
contrasting profiles (Figure 7). The H \& K lines are broad lines without a hint
of a sharp emission component but with two (may be even more) sharp absorption
features. The IR lines and forbidden  lines appear as sharp lines 
without a broad
emission component.
The absence of the broad emission in the IR lines is consistent
with the flux predicted from the branching ratio and the
flux in the H \& K lines.
 The broad emission of the H \& K lines is fairly well described by a
single Gaussian with a mean FWHM of 220 km s$^{-1}$ at a mean velocity of
$-4$ km s$^{-1}$. These parameters are quite similar to those
derived from the Gaussian fit to the Na D lines.

There is a distinct difference in the
Ca\,{\sc ii} H and K
  broad emission profile in the 2003 and the 1995-1996 declines
(Figure 8).
 The 2003 profile is well fitted by a single Gaussian.
 The 1995-1996
profiles require two Gaussians with a separation of 181 km s$^{-1}$ and a red Gaussian
about 65 \% stronger than the blue one. The blue Gaussian has the larger FWHM
(204 km s$^{-1}$ to 167 km s$^{-1}$). The
Gaussian fits for Ca\,{\sc ii} H and K are similar to fits to the
Na D lines for the 1995-1996 and the 2003 profiles.
 These similarities suggest that the emitting regions for the Na
atoms and the Ca$^+$ ions are
very closely related, if not identical, both in 1995-1996 and 2003.
However, the common emitting regions were differently arranged with respect to
 the stellar radial velocity at the two minima.

\begin{figure}
\epsfxsize=8truecm
\epsffile{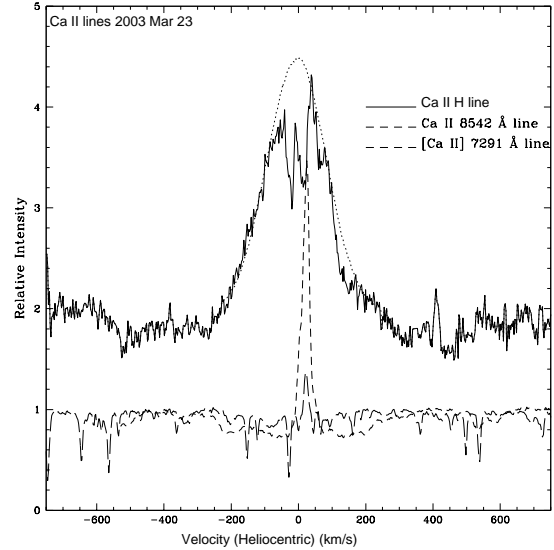}
\caption{Profiles of three Ca\,{\sc ii} lines in 2003: the resonance  H line,
the IR line at  8542 \AA, and the  [Ca\,{\sc ii}] line at 7291 \AA.
Telluric H$_2$O lines contaminate the region around the
[Ca\,{\sc ii}] line. }
\end{figure}

\begin{figure}
\epsfxsize=8truecm
\epsffile{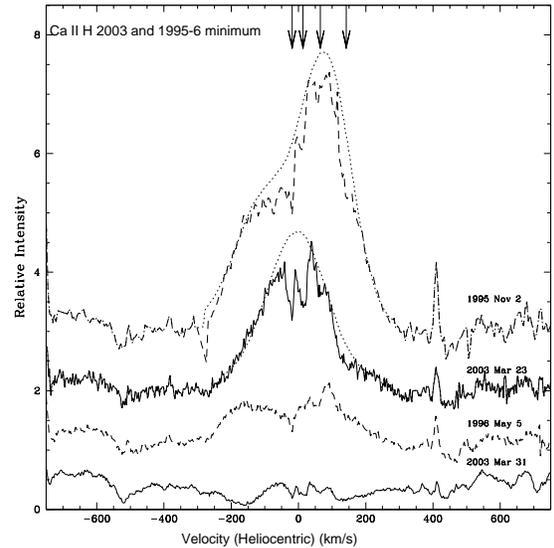}
\caption{Profiles of Ca\,{\sc ii} lines  H line in 2003 and 1995 minimum. Gaussian fits
to the profiles of 2003 March 23 and 1995 November 2 are also shown (dotted lines).The
arrows indicate the positions of superposed absorption components.}
\end{figure}

Of interest in the H \& K profiles are the absorption
features at $-21$ km s$^{-1}$, $+16$ km s$^{-1}$, and $+65$ km s$^{-1}$,
and the absence of a sharp emission component, the analogue of the so
prominent sharp emission component to the Na D lines. The $-21$ km s$^{-1}$
absorption matches the velocity of the narrow Na D absorption
seen at maximum light (and in decline) and attributed to
interstellar gas (Keenan \& Greenstein 1963; Payne-Gaposhkin 1963; Rao \&
Lambert 1997). The $+65$ km s$^{-1}$ component has no counterpart
in other broad emission lines, but variable absorption redshifted relative
to the mean stellar velocity is seen in spectra from maximum light
(see below). That this redshifted gas is not seen in Na D is probably
because  Na atoms
are singly-ionized in this infalling
gas.
The $+16$ km s$^{-1}$ component with its blueshift of about 6 km s$^{-1}$
has a velocity similar to that of the sharp IR and the 7291 \AA\ and
7323 \AA\ lines (Figure 7).
 (The peak sharp absorption velocity in the H \& K lines may be
shifted to the blue by a few km s$^{-1}$ relative to the IR and forbidden
line velocities and the typical ionized `metal' line.)

 Sharp line emission in H \& K has been
replaced by absorption.
If the sharp line region were optically thin to the Ca\,{\sc ii}
lines, the flux in the H \& K sharp lines estmated from  
the branching ratio and the flux in the IR lines would be
several times the actual flux in the broad H and K lines, but observations
show, as in 1995-1996, that the sharp emission lines are absent.
Absorption seen as several apparently discrete components is
in part presumably due to optically thick gas projected in front of part or all of the
broad line emitting region.
% That sharp emission expected in the H \& K
%lines has been replaced by absorption implies that the broad line overlies
%the region emitting the sharp lines and is optically thick to the
%resonance photons.

A likely principal reason for the absence of the sharp
lines is that the emitting
layers are optically thick to the resonance lines, as suggested by the
following argument.
          If the
       IR triplet lines are optically thin, the intensity ratios expected
      for 8498 \AA\ : 8542 \AA\ : 8662 \AA\ are 1:9:5.
      Observations at the  1995-1996 minimum show roughly equal fluxes
for the three lines, i.e., the emitting gas is not optically thin.
The optical depth in the H \& K lines has certainly to be several
times higher than in the IR lines.
         The observed flux ratio of the [Ca\,{\sc ii}] lines to the IR triplet
      lines is estimated to be 0.03.
         The calculations of Ferland and Persson  (1989) suggest that for an
electron temperature of T$_{\rm e}$
      between
      3000 to 10000 K that the observed flux ratio of forbidden to IR triplet
      lines is found for electron densities of  N$_{\rm e}  \sim  10^{10\pm1}$ cm$^{-3}$,
a value consistent with other estimates for the 1995-1996 minimum (Rao et al. 1999).
Their calculations indicate that the observed fluxes and the inferred electron
densities imply optical depths in the H \& K lines of $\tau \sim 10^4$. Thus, we
suppose that suppression of sharp emission in the resonance lines is possible.

\subsubsection{[N\,{\sc ii}]  lines}

The [N\,{\sc ii}] line at 6583 \AA\ is weakly present in 2003. The maximum
intensity  is about 10\% of the local continuum in the 2003 March 8 and
13 spectra.  (The  weaker
 6548 \AA\ and 5754 \AA\ lines  are not detected.) The profile
appears to be very similar to the well determined profile
from the 1995-1996 minimum: broad with  a central
minimum between two peaks of equal intensity.
This profile differs from that presented by other broad emission lines.
 With the accurate rest
wavelength (6583.454 \AA)  determined by  Spyromilio (1995) and
Dopita \& Hua  (1997),
the centre of the line in 2003 is at a velocity of
$-6\pm4$ km s$^{-1}$. The
velocity range at the base extends from -163 to 140 km s$^{-1}$.
The flux in the line is about
1.2$\pm0.2 \times 10^{-14}$ erg cm$^{-2}$ s$^{-1}$ on 2003 March 8 and 13
whereas the flux on 1996 February 6 is
1.7$\pm0.2 \times 10^{-14}$ erg cm$^{-2}$ s$^{-1}$.
Differences in profile and flux betweem 1996 and 2003 are not be significant.

%\begin{figure*}
%\epsfxsize=18truecm
%\epsffile{rcrb2003fig4.eps}
%\caption{[N II] line $\lambda$ 6583 in 2003 and 1995 - 96 minimum. }
%\end{figure*}.

\subsection{The Sharp emission lines}

Available data suggest that sharp emission lines
of abundant singly-ionized and neutral metals, are seen in all
declines of all RCBs from soon after onset of a decline through minimum light to near
full recovery to maximum light. The lines generally show a slight
blueshift with respect to the mean velocity of the star as measured
from absorption lines at maximum light. First, we comment on these
sharp lines as seen in the 2003 spectra and then report on a first
for R CrB -
sharp emission lines of the C$_2$ molecule.

\subsubsection{Sharp emission lines -- atoms}

  Our 2003 spectra are rich in sharp emission lines of
 Sc\,{\sc ii}, Ti\,{\sc ii}, Fe\,{\sc i},
 Fe\,{\sc ii}, Y\,{\sc ii}, Ba\,{\sc ii}
and other species with a level of excitation similar to
that in 1995-1996. The lines in 2003 are broader (FWHM $\simeq
20$ km s$^{-1}$) than in 1995-1996 (FWHM $\simeq 15$ km s$^{-1}$).
Radial velocities of
 these lines are shown in Table 2. It will
be seen that the  blueshift  relative to the
mean stellar velocity at maximum light
 of $+$22.5 km s$^{-1}$ is absent or very small
for our 2003 spectra.
Except for a suggestion that the velocity may have been slightly different
on 2003 March 8, the sharp lines do not show a velocity variation over this
interval of less than a month. This lack of a velocity variation
is consistent with a near-constant velocity for these lines over the entire
duration of the 1995-1996 minimum.

The sharp  emission lines in the 1995-1996 spectra often but not always
 showed velocity
structure suggestive of a blend of two or three components. In 2003, the
line profiles  arise from a single component with a velocity equal to
that of the strongest component seen in 1995-1996. Such profiles were seen
on occasion in 1995-1996.
%Red-shifted absorption is seen in the red wings of the IR Ca\,{\sc ii} lines
%(Figure 8) at a radial velocity of 40.8 km s$^{-1}$.
       The mean level of excitation decreased between
March 8
 and March 13;  the Fe\,{\sc ii} lines  weakened and Fe\,{\sc i}  lines
 got stronger relative to the continum. which, as judged by the V
magnitude, was approximately constant in this six day period.

%       The V mag of the 2003 March 23 and the 1995 Oct 18 are about the same
% only the 2003 spectrum is on the recovery light where as the 1995 spectrum
% was obtained on the decent to minimum. The absorption spectrum relative to
% the continum (line depths etc.) seem to match very well . However the sharp
% emissions so difference eg. in the level of excitation (lower in the 2003
% spectrum ) and line widths eg. the sharp lines of Na I D are slightly broader
% in 2003 minimum. The Ca II triplet line $\lambda$ 8542 much broader
% in 1995 Oct 18 th than on 2003 March 23 - corespondingly the superposed
% absorption component widths are also broder in 1995 spectrum relative to
% 2003 spectrum reflecting the fact that the IR triplet lines are flourescent
% lines .

%\begin{figure*}
%\epsfxsize=18truecm
%\epsffile{plot14rcrb03sefeb28.ps}
%\caption{Profiles of sharp emission line$\lambda$8542 in 2003 mar 31 and light
% maximum}
%\end{figure*}.

A few forbidden
transitions are seen as sharp emission lines:
[Ca\,{\sc ii}] at 7291 \AA\ and 7323 \AA, [O\,{\sc i}] at  6300 \AA,
 and [Fe\,{\sc ii}] at  7155 \AA\
 in the spectral
 region covered in the HET spectra at minimum.
 The [Ca\,{\sc ii}]
 lines were also present in the spectra obtained later in March on  the
 recovery from minimum.
The [C\,{\sc i}] 9850 \AA\ line
 is  present in the spectrum of March
 23.
% The [Fe\,{\sc ii}]
% lines had
% disappeared by the time of our recovery spectra. These forbidden
%lines were
%seen during the  1995-96 minimum. The electron density for the emitting
%gas inferred from the lines
% in the 1995-96  spectra  of 10$^{ 7}$ cm$^{-3}$ might still
% be applicable in 2003.

    The emission lines in 2003  are of
higher  relative intensity at a given V magnitude than in 1995-96 minimum.
 Figure 9   shows the spectrum in the region of the Si\,{\sc ii} 6371 \AA\
line. The 2003 March 8 spectrum (V = 12.8) is compared with
 1995 November 2 (V =12.2)
 on the descent to
 light minimum
and at light minimum on
1996 February 6 (V=13.5)
 when the lines are in absorption.
 Comparison of spectra at the same V magnitude during the recovery to maximum
light from the minimum gives the same result -- see the spectra of
2003 March 23 and 1996 May 5 in Figures 9 and 10.
These differences are not unexpected because our  coverage of the sharp
lines through 1995-1996 minimum showed that a line's flux changed,
presumably due to varying obscuration of the emission region by dust.

\begin{figure}
\epsfxsize=8truecm
\epsffile{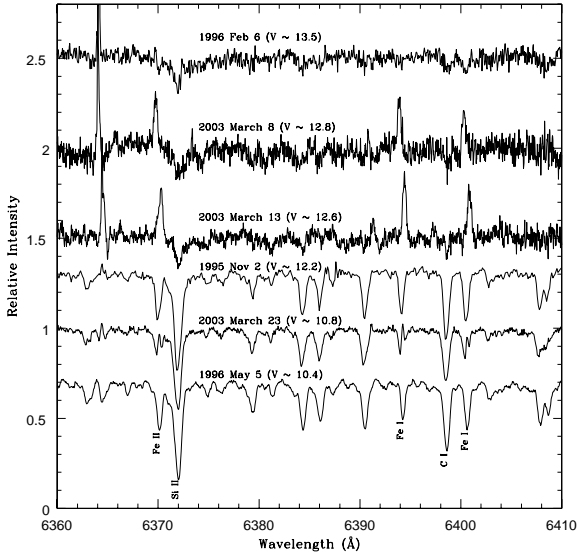}
\caption{Sharp emission lines in the 6370 \AA\ region in the
1995-1996  and 2003 minima.}
\end{figure}

\begin{figure}
\epsfxsize=8truecm
\epsffile{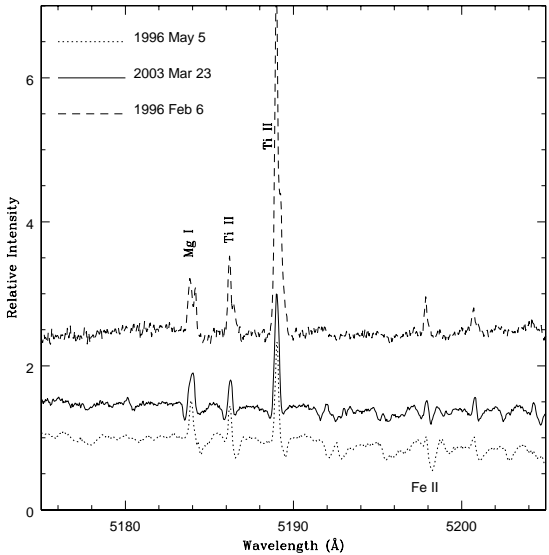}
\caption{Sharp emission lines near 5190 \AA\ at the 1995-1996 and 2003 minima.}
\end{figure}

\subsubsection{Sharp emission lines - the C$_2$ molecule}

The 2003 minimum spectra for March 8 and 13
  show C$_2$ Swan bands  in emission.
 The observed spectral region included the 1-2, 0-2, and 1-3
bands.
 The rotational structure is well resolved but
at the 1995-1996 minimum  the lines were so broadened
that rotational structure was not resolved (Rao et al. 1999). Such a change
from sharp to broad Swan lines has been seen previously in V854 Cen:
the 1998 minimum of V854 Cen showed sharp Swan lines 
but the lines were broad at the 1992 minimum (Rao \&
Lambert 2000).
% This behaviour
% is 
%a  clue to the location of the molecular gas.

\subsubsection{C\,{\sc i} emission lines}

A characteristic of the sharp `metal' lines is the near constancy of their
velocity (and profiles) during the decline and recovery from
minimum light. The C\,{\sc i} emission lines in contrast appear
with a variable velocity.
The  lines were in emission on 2003 March 8 (Figure 11)  with
a  velocity of  15 km s$^{-1}$.
 The emission
moved redward achieving a velocity
of 37 km s$^{-1}$ on March 13.
The observation that the velocity of the C\,{\sc i} emission migrates from
13 to 37 km s$^{-1}$ but the low excitation sharp metal emissions
are stationary suggests a distinctly different character to the
high excitation emissions.

\begin{figure}
\epsfxsize=8truecm
\epsffile{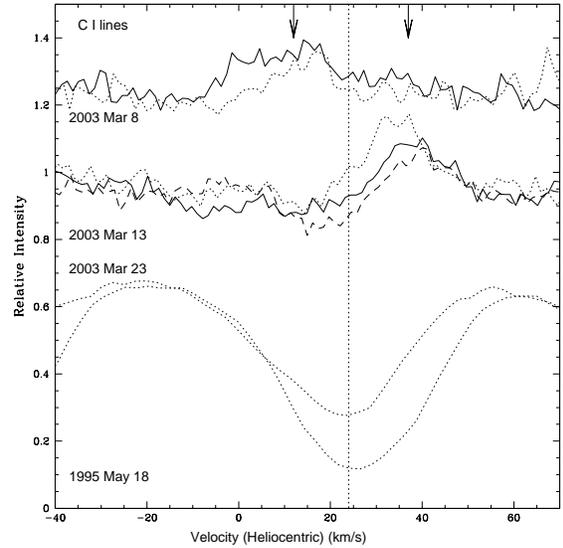}
\caption{Emission in the C\,{\sc i} lines at 7111, 7113, and 7115 \AA\
lines at the 2003 minimum. The velocity scale is set for the
7113 \AA\ line with the vertical line representing the average heliocentric
velocity of the star at maximum light.
 The solid line shows the 7113 \AA\ emission profile for 2003 March
8 (top) and 2003 March 13 (middle) and the absorption profile for
2003 March 23.
 The dotted line (top and middle) shows the 7111 \AA\ profile
and the dashed line (middle)  shows the 7115 \AA\
profile  for the same velocity zero point as the 7113 \AA\ line. Emission
on March 8 is centred at $+13$ km s$^{-1}$ and on March 13 at $+37$ km s$^{-1}$
(see arrows at the top of the figure).}
\end{figure}

In the spectra obtained after March 8, the C\,{\sc i} lines are
in absorption but weaker than at maximum light with
an apparent filling in by emission.
For example,
 the  6828 \AA\ line 
not only shows movement of the emission
 peak from blue to red from 2003 March 8 to
13, but also shows  variable emission in the absorption core  between 2003
March 13 and March 31 with a tendency to oscillate between a velocity
of about $+11$ and $+30$ km s$^{-1}$ (Figure 12).

\begin{figure}
\epsfxsize=8truecm
\epsffile{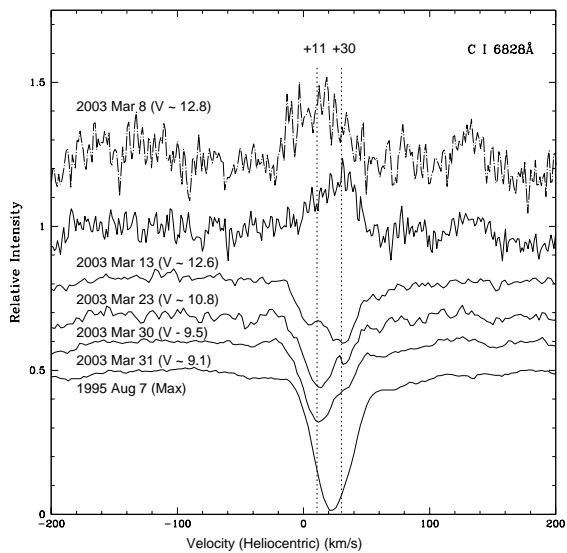}
\caption{Profiles of the C\,{\sc i} 6828 \AA\ line
on 2003 March 8, 13, 23, 30, and 31 along with the profile
at maximum light from 1995 August 7. The V magnitudes of R CrB at the time of
observation are given on the figure.}
\end{figure}

Pure emission at minimum light, as in Figure 11 and 12, was not seen in the
1995-1996 but our sampling then was less than optimal; minimum
light coincided with R CrB's passage behind the Sun and on its
emergence, when the star was recovering to maximum, our
data were sparsely spaced in time. Nonetheless, the C\,{\sc i}
profiles were variable to a similar extent and on a comparable
timescale in 1995-1996 and 2003. This is shown by Figure 13.
High excitation lines of other atoms show similar short period changes in
their line profiles.

\begin{figure}
\epsfxsize=8truecm
\epsffile{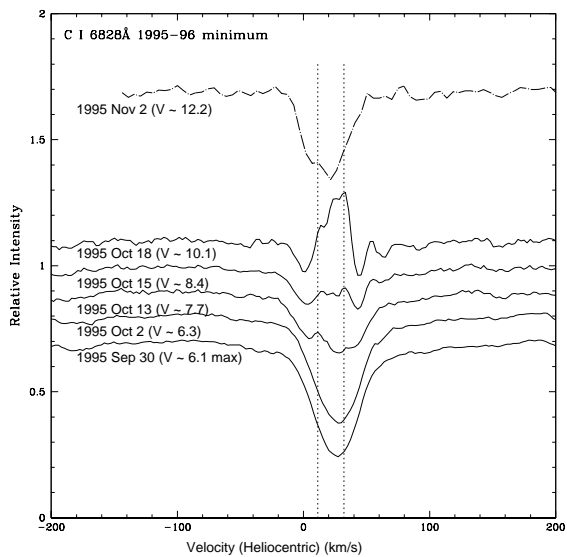}
\caption{Profiles of the C\,{\sc i}  6828 \AA\ line
on 1995 September 30, October 2, 13, 15, 18, and November 2.
The V magnitudes of R CrB at the time of the observations are given
on the figure.  The vertical dashed lines indicate
radial velocities  of 11 and 31 km s$^{-1}$.}
\end{figure}

%\begin{figure*}
%\epsfxsize=18truecm
%\epsffile{plot13rcrb03sefeb28.ps}
%\caption{Profiles of Si\,{\sc ii} 6371 \AA\ line
%on 2003 March 8, 13 ,23 ,30, and 31. Dashed vertical lines indicate
%radial velocity of 10 and 37 Km s$^{-1}$. }
%\end{figure*}.

%Note that the neighbouring 6369 \AA\ Fe\,{\sc ii}
%does not show this emission bump. There  appears  to be
%an emission bump present in H $\alpha$ profile.
% at about 16 Km s$^{-1}$
%on 2003 Mar 8 (-may be no emission or weak emission in the noisey
%profile on 2003 Mar 13 - figure).

This transient emission in  C\,{\sc i} and other 
  lines in the
 1995-1996
minimum was interpreted as a signature of shock propagation
related to the photospheric pulsation (Rao et al. 1999), as suggested in the model proposed
by Woitke et al. (1996). The repeated occurrence of C\,{\sc i} emission
migrating from blue to red in a period of about 5 days suggests
such a propagating pulsation-related shock with molecules forming
in cooled gas behind the shock (Woitke et al. 1996).
The shock cannot be a normal feature of the photosphere because
the C\,{\sc i} emission, if seen without obscuration by dust, would
be very intense and a strking feature at maximum light, but
all of our spectra taken at maximum light show pure absorption
profiles with no hint of overlying emission. The emission may
come from a  disturbed  photospheric
region associated with the region of dust production.
More likely, the emission may be produced above the photosphere
where a shock also leads to dust.

\section{Absorption lines of the  C$_2$ Phillips system}

One supposes that dust formation is preceded and accompanied by
molecule formation. The C$_2$ molecule is expected to be
an abundant molecule, and, as noted above, sharp emission from these
molecules is seen in the Swan bands.
The lower level of these bands is not the molecule's
ground state but connected to it by a forbidden transition.
C$_2$ absorption may  more likely to be detected using the Phillips
system, an electronic
transition from the ground state. Absorption in the Phillips
bands was previously seen in V854 Cen during its 1998 minimum
(Rao \& Lambert 2000).
                                                                                 
Examination of our 2003 March 23, 30, and 31 spectra of R CrB
showed the presence of weak absorption lines corresponding to the
lines of
the  2-0 Phillips band
(Figure 14) in the March 23 but not the March 30 and 31 spectra.
 The HET spectra
 obtained earlier during the minimum did not cover  the spectral region of
either the 2-0 or the 3-0 Phillips bands.
The C$_2$ Phillips lines were not seen in absorption (or emission) in the
1995-1996 minimum for which our spectral coverage was
extensive. This  absence of the Phillips
lines is very likely explained by the fact that C$_2$ Swan
emission lines in 1995-1996 were broad and thus one may
suppose that attendant absorption lines would also have
been broad and so undetectable.

\begin{figure}
\epsfxsize=8truecm
\epsffile{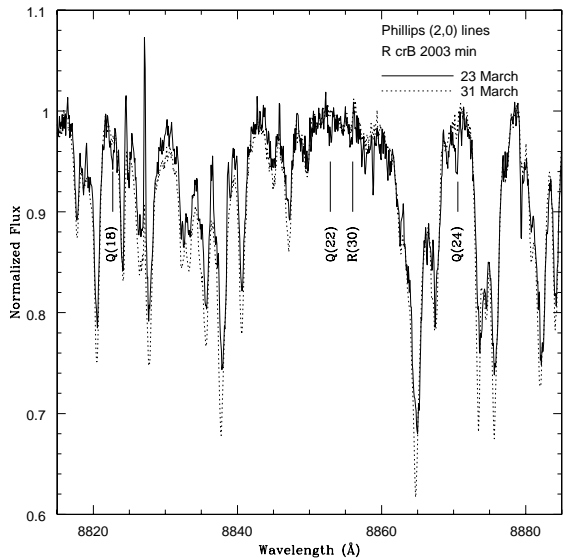}
\caption{Sharp absorption lines of the C$_2$ Phillips (2-0) band on 2003 March 23. The
C$_2$ lines are not present in the 2003 March 31 spectrum.}
\end{figure}
                                                                                 
A Boltzmann plot constructed from the equivalent widths of seven  2-0 lines suggests
 a rotational excitation temperature of
 T$_{\rm rot} = 1377\pm150$ K (Figure 15), assuming the lines are
unsaturated.
 The mean radial velocity of the  lines is $+$10
 km s$^{-1}$  equivalent to a blue shift of 13 km s$^{-1}$ with respect to the
photosphere. For comparison, we note the  
excitation temperature for the C$_2$ molecules at the
1998 minimum of V854 Cen was 1100 K.
%suggests an association of shocked region and the cool gas.

\begin{figure}
\epsfxsize=8truecm
\epsffile{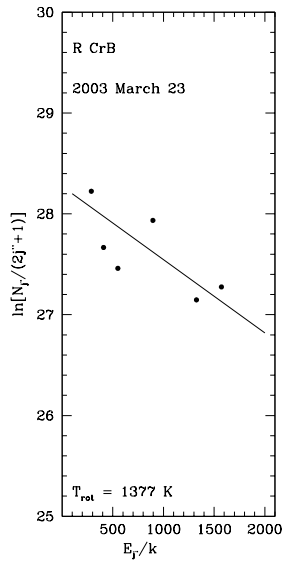}
\caption{Boltzmann plot for the   C$_2$ Phillips 2-0 absorption lines from
the  2003 March 23 spectrum.}
\end{figure}

    The C$_2$ molecules are likely to reside close to the freshly formed
dust, probably below the dust cloud. On March 23 when the sharp Phillips 
lines were detected in absorption, the dust cloud was not providing
veiling of the photospheric lines (Figure 9).
Detections of 
C$_2$ Swan bands in emission on March 8 and 13 were made when veiling
of photospheric lines was evident (Figure 9) yet the C$_2$ lines
were sharp. Then, the C$_2$ molecules can not have been viewed through
the parts of the dust cloud projected on to the photosphere. The line
of sight to the molecules must have been dust free or relatively so.

\section{Veiling of the photospheric lines}

The spectra of 2003 March 8 and 13  are almost completely devoid
of absorption lines, even the strongest lines such as the
Si\,{\sc ii} 6347 \AA\ and 6371 \AA\ lines (Figure 9) are
greatly weakened and possibly broadened. By March 23 when our next
spectrum was obtained, the absorption spectrum had largely
returned to its appearance at maximum light.
A similar behaviour of absorption lines  was
extensively described for the 1995-1996 minimum by
Rao et al. (1999). Figure 9 shows three spectra from that
minimum including the one for 1996 February 6. Scattering
of photospheric light passing through the dust cloud
Doppler-broadens the light and washes out the absorption
lines.

\section{An indicator of a stellar wind}

The spectrum of R CrB is so rich in novelties that some
may be missed by even serious inspection. In examining the
new spectra, we noticed that the
O\,{\sc i} triplet lines at 7771-5 \AA\  show extended
blue wings  on the profiles obtained on 2003 March 23,  30, and 31 (Figure 16).
This extension is clearly revealed  by comparison of the blue wing of the
7772 \AA\ line and the red wing of the 7775 \AA\ line and by contrast
with the O\,{\sc i} profiles for $\gamma$ Cyg where the lines are of
a similar strength.
The extended blue wing is seen in hindsight on all spectra taken with
R CrB at maximum light as well as those from the 1995-1996
decline.
The wings  extending to velocities of about -130 km s$^{-1}$ from line
centre are suggestive of a strong wind.
A wind velocity of over 130 km s$^{-1}$ exceeds the
predicted
escape velocity of 30--70 km s$^{-1}$ (Rao \& Lambert 1997).
                                                                                 
A wind was previously suggested by
observations of the He\,{\sc i} 10830 \AA\ line at maximum light showing
blueshifted absorption at a velocity of about 200 km s$^{-1}$
in spectra acquired at or near maximum and spanning
more than two decades 
(Querci \& Querci 1978; Zirin 1982; Clayton, Geballe \& Bianchi 2003).
Some spectra have shown also redshifted emission suggesting a
P Cygni-like profile. A point for He\,{\sc i} is added to Figure 16.

\begin{figure}
\epsfxsize=8truecm
\epsffile{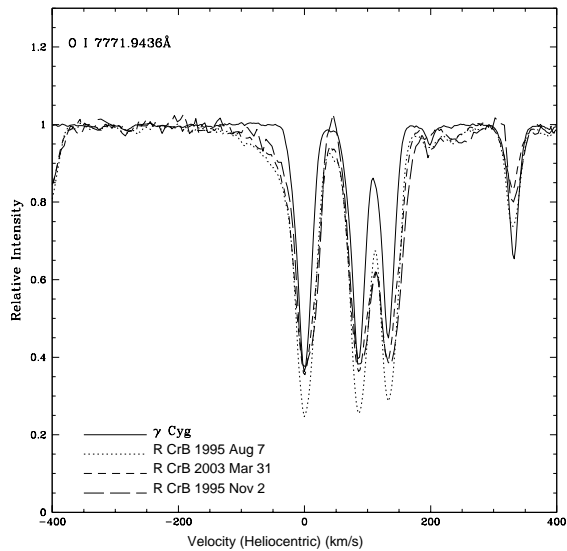}
\caption{Profiles of the O\,{\sc i}  7772 \AA\  triplet on
2003 March 23 and 31 and
at maximum light on 1995 August 7  compared with the profile in $\gamma$ Cyg.
The zero point of the velocity scale is set for the bluemost
line of the triplet. }
\end{figure}
                                                                                 
Strength of the wind as seen in O\,{\sc i} appears to be nearly
constant and little, if at all, changed from maximum to the
faintest magnitudes at which the absorption spectrum is unaffected
by veiling. Figure 17 shows seven spectra from
1995 May 18 to 1995 September 30 in which six show
essentially identical profiles with extended blue wings. The
seventh spectrum taken on 1995 September 30 was taken
one day prior to the onset of the 1995-1996 decline and
its more extended blue wing may have been a harbinger of the
decline.
                                                                                 
In contrast to the O\,{\sc i} profiles, there is appreciable
variation in the blue wing of the Al\,{\sc i} 3944 \AA\
resonance line. (Similar variations are present in the
Na D lines (Rao \& Lambert 1997 and Figure 19) but their 
definition is compromised by the
presence of the blue-shifted interstellar Na D lines
and telluric H$_2$O lines.)
Figure 17 shows the spectrum around the Al\,{\sc i} line
for the same dates between 1995 May and September. Large
variations in the absorption to the blue of the photospheric
Al\,{\sc i} line are seen. Several spectra show a `cloud'
at $-40$ km s$^{-1}$. Others show an extended blue wing
out to about $-50$ km s$^{-1}$. These variations on the
Al\,{\sc i} profile stand in stark contrast to the lack of
profile variations in the photospheric line shown at
$-130$ km s$^{-1}$.

\begin{figure}
\epsfxsize=8truecm
\epsffile{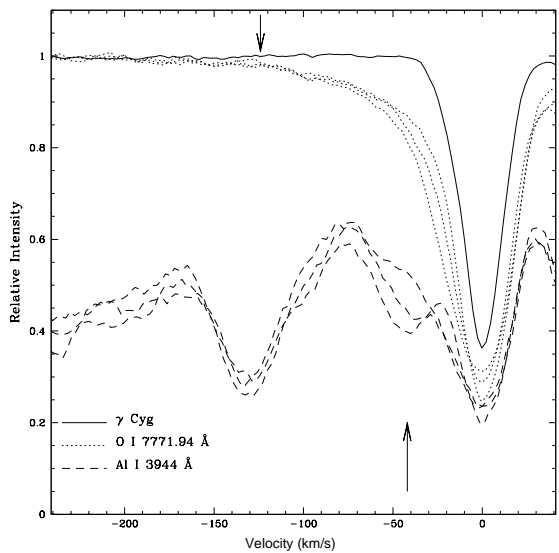}
\caption{Seven spectra of the O\,{\sc i} 7772 \AA\ (top)
and the Al\,{\sc i} 3944 \AA\ line of R CrB at maximum light
from 1995 May 18 and September 30. The O\,{\sc i} line from
$\gamma$ Cyg is  shown (solid line). The velocity zero point is set at
the centres of the O\,{\sc i} and Al\,{\sc i} lines.}
\end{figure}

Inspection of spectra show that the
inferred wind velocity decreases
with excitation potential of the line:
Si\,{\sc ii} at 6347 \AA\ and 8.3 eV shows 82 km s$^{-1}$,
C\,{\sc i} at 6828 \AA\ and 8.5 eV shows 107 km s$^{-1}$, and
9112 \AA\ and 7.5 eV shows 100 km s$^{-1}$,
Fe \,{\sc ii} at 6369 \AA\ and 2.9 eV shows 63 km s$^{-1}$  
, Fe \,{\sc i} at  7511 \AA\ and
4.1 e.v shows 52 km s$^{-1}$, and Al \,{\sc i} (and Na \,{\sc i} D
lines) of 0.0 e.v showing 44 km s$^{-1}$.
Clayton, Geballe \& Bianchi's (2003) reported wind velocities of 200 to 240 km s$^{-1}$
at maximum from the  He\,{\sc i} 10830 \AA\ line,  apparently
extend the velocity -
excitation relation shown in Figure 18.
In addition to the outflowing gas comprising the stellar wind, there is
evidence for infalling gas of variable strength at maximum light, as revealed by the 
Ca\,{\sc ii} H \& K and IR lines (Figures 19 and 20). This gas is not
seen in the Na D lines; the red wing of the Na D lines is unchanged
during the Ca\,{\sc ii} variations (Figure 19).

\begin{figure}
\epsfxsize=8truecm
\epsffile{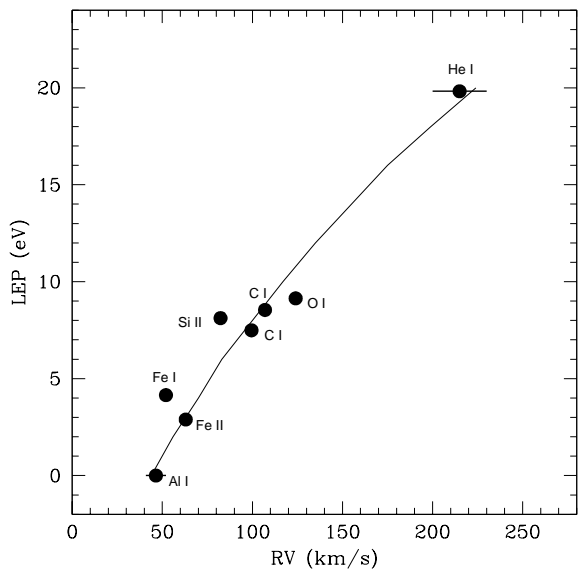}
\caption{Wind velocity versus lower excitation potential
of the line for spectra at maximum light from 1995 May to
September. The data point for He\,{\sc i} is from
Clayton et al. (2003).}
\end{figure}
                                                                                 
\begin{figure}
\epsfxsize=8truecm
\epsffile{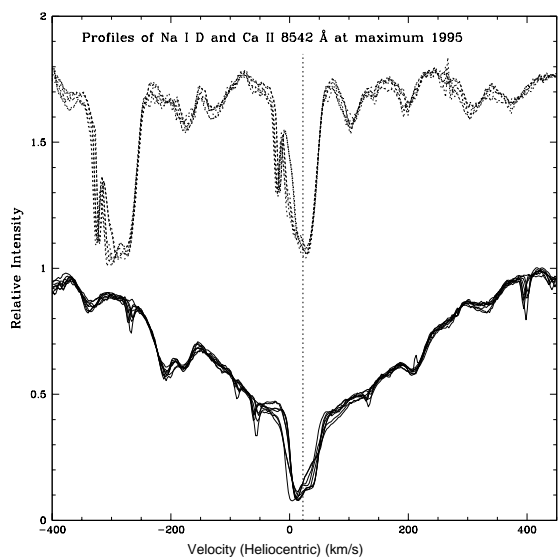}
\caption{Presence of stellar wind in Na \,{\sc i} D (top) and Ca\,{\sc ii}
8542 \AA\ (bottom) lines from spectra at maximum light from 1995 May to
September. Note the absence of redward extensions in Na \,{\sc i} D 
lines where as Ca\,{\sc ii} 8542 \AA\ line shows both red and blue
extensions to the profile (variable). The vertical dashed line refers to the
average radial velocity of the star. The profiles are normalised to
the stellar velocity.}
\end{figure}

\begin{figure}
\epsfxsize=8truecm
\epsffile{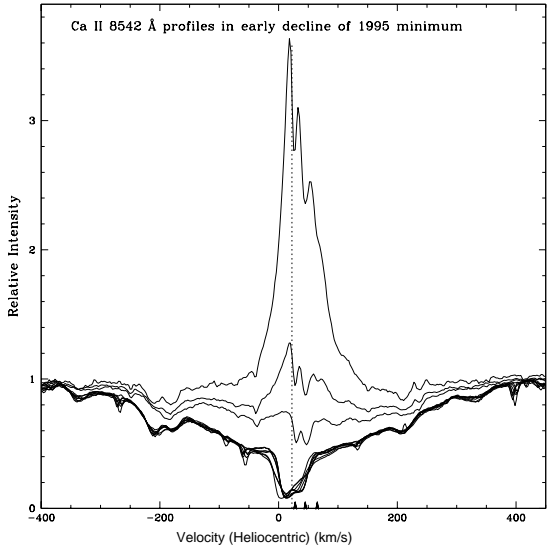}
\caption{ Ca\,{\sc ii}
8542 \AA\  line during early decline of 1995-1996 minimum. 
The lowest set of profile refer to maximum light
The upper profiles show the change the minimum progress from 6 to
10.2 magnitude. Note the presence of redward superposed absorption 
components at +28, +45 and +66 km s$^{-1}$. 
The vertical dashed line refers to the
average radial velocity of the star of 22.5 km s$^{-1}$.}
\end{figure}
 
\section{Discussion}

Even a qualitative interpretation of the spectrum of R CrB in decline
must today be circumscribed with qualifications and admissions of
ignorance. Our comparison of spectra from the 1995-1996 and 2003
declines adds new information whose full significance may become
clearer with observations from future declines of R CrB. 
Here, we restrict comments to general features of the absorption and
emission spectrum of the star.

Our discovery of extended blue wings to strong lines observed
at maximum light suggests the presence of a permanent wind off the
surface of R CrB (and the other RCBs for which we have suitable
spectra). The wind velocity increases with a line's excitation
potential. Variable non-photospheric Doppler-shifted absorption is
seen in several resonance lines: blueshifted absorption in Na D and
Al\,{\sc i} and blueshifted and redshifted absorption in the Ca\,{\sc ii}
H \& K and IR lines. These variable components which suggest
a circulation of gas  may be linked to the photospheric pulsations.

      Evidence for a wind off R CrB and other RCBs was earlier provided
by Clayton et al. (2003) from observations of the He\,{\sc i} 10830 \AA\
line showing blue-shifted absorption with or without accompanying emission.
Clayton et al,'s modelling of 
the profiles suggested  that the line is formed by collisional excitation
and that acceleration of the wind to its possibly terminal velocity (indicated by the
10830 \AA\ profile) is steep.
 Such conditions in the wind are
likely conducive to the absorption of O\,{\sc i} 7772 \AA\ by the
wind.
 The base of the wind must cover large fractions of the
photospheric area in order that the blue wing of a strong
line vary little, if at all, at maximum light.

  One interpretation of the extended blue wings is that the wind
begins at the top of the photosphere  and
increases in velocity and excitation with height above its
base. Heating of the wind may be by deposition of mechanical
energy (sound and/or hydromagnetic waves);
the photospheric absorption lines have a width indicating
mass motions with a velocity exceeding the
local sound speed of about 5 km s$^{-1}$ (Rao \& Lambert 1997).

A decline occurs when a soot cloud  begins to form along the
line of sight to the star. The trigger for dust formation may
be a pulsation-induced outwardly propagating shock (Woitke
et al. 1996). Calculations indicate that temperatures
behind a shock can drop to below 1500 K at densities sufficiently
high for carbon atoms to stick to provide an obscuring layer
of soot. The shock and soot may initially form over a small
part of the visible photosphere but lateral expansion of the soot
cloud will result in obscuration of the entire earth facing hemisphere
 of the photosphere. As soot
forms and spreads, the sharp emission lines are revealed.
While the cloud may be a local phenomenon, the regions emitting the
sharp lines would appear to be more widespread both in terms of spatial
coverage and in height above the photosphere (Rao et al. 1999).
The intensity of the sharp lines in contrast to the intensities of the
broad lines 
clearly declines with increasing obscuration
of the star; The broad line region must be even more extended than
the sharpline region. 
When the photospheric 
absorption lines are `veiled' by a thick cloud of soot, the sharp
lines, although diminished in flux, remain sharp. Therefore, the
region emitting the lines extends off the photosphere and beyond
the boundaries of the soot
cloud.\footnote{ Rao, Reddy \&
Lambert (2004) speculated that RCBs may posses a dusty equatorial torus
and polar winds. The polar winds provided the sharp emission lines
 from heights relatively close to the star. The sharp line's 
velocity (relative to the mean stellar velocity) depends on the circulation
of the polar axis and the extant to which the fresh cloud blocks the 
approaching and receding polar winds.}

   The several types of broad emission lines present interpretative
challenges. The He\,{\sc i} lines with their quasi-parabolic
profiles and a mean variable velocity close to the photospheric
mean velocity may be associated with the wind: the velocity offsets relative to
the mean velocity was $-60$ km s$^{-1}$ in 2003 and $-30$ km s$^{-1}$ in 1995-1996.
 If the emitting region
of the wind has a radius at least a few times the stellar radius
and is approximately uniform -- back and front, the broad line will have a
rather smooth profile centred on the mean stellar velocity with  a width
that is about twice the wind's velocity for the helium emitting
layers. The lines' blueshifts may result from occultation of the
receding wind behind the star. Weak redshifted emission also suggests
that occultation is occurring.
 Assuming that the velocity of the blueshifted absorption
in the He\,{\sc i} 10830 \AA\ line reported by Clayton et al.
(2003) is representative of the He emitting region, the 
observed FWHM of our He\,{\sc i}
lines is accounted for in a qualitative sense.
Departures from a smoothly distributed wind or variations in
helium line excitation will result in profile variations.\footnote{One
expects a drift velocity between dust grains and gas atoms. This
velocity offers the potential for either  collisions
between dust grains and helium atoms or collisions between
helium atoms and electrons reflected off grains to lead to
excitation of helium atoms and helium line emission.}

   The broad emission profiles of the Na D and Ca\,{\sc ii} H and K lines
are  similar for a particular decline but this profile in 1995-1996 differed
clearly from that in 2003: a single Gaussian  fits 
the 2003 profile but two well-separated Gaussians are required to
fit the 1995-1996 profiles. Yet, the mean velocity of the broad
emission at both declines was similar and close to the mean
stellar velocity. Additionally, the line fluxes appear to
be little changed during and between declines suggesting that the
bulk of the emitting region is unobscured by
 the fresh soot cloud.\footnote{The Na D broad emission
from
V854 Cen in a deep decline was unpolarized when the continuum was
polarized showing that the emission in that case
was also unobscured by the dust cloud responsible
for that decline (Whitney et al. 1992).}

  In contrast to the broad helium emission lines, the
velocity width of the Na and Ca$^+$ emission is much greater than wind
velocity of Na and Ca$^+$ atoms, as
indicated by the blue extensions to the wings of the
photospheric Na D and Ca H \& K  absorption lines.
 Although the 2003 Na D and Ca H \& profiles are similar to those
of the helium emission (suggesting a  common origin for He, Na, and
Ca lines in the outer
reaches of the wind),
  colocation of He and Na atoms and Ca$^+$ ions
is not supported by
the profiles in 1995-1996 of Na D and Ca H \& K lines which differed
dramatically from the
the He\,{\sc i} emission
profiles: the Na D and Ca H \& K lines appeared as two well separated
Gaussians
but the He lines were fit by a single Gaussian.
Although the Na D and Ca H \& K profiles  changed
so greatly between the two declines,
 the central emission velocity was unchanged and at a value only
slightly blueshifted from the mean stellar velocity. Apparently,
the emitting Na atoms and Ca$^+$ ions were approximately
 symmetrically distributed
about the star but with  different distributions on the two occasions; perhaps, for example, the emitting gas is in a variable with polar symmetry.
 Perhaps also, the dissimilar but symmetric
distributions from 1995-1996 and 2003 are merely fortuitous and future observations of
the Na D and Ca H \& K lines will show a wide variety of profiles
with  mean velocities differing
appreciably from the mean stellar velocity. Then, firmer clues to the
geometry of the broad emission line regions should be
provided.

Indeed, the unresolved key questions about the various emission and
absorption features seen at minimum light may be thought to concern the
geometry of R CrB's outer
atmosphere. To conclude, we list the questions and comment very
briefly on some geometric questions.

Does the wind have a latitude dependence? Is this variation slight or
extreme? We noted above our earlier suggestion that the wind off a RCB may
be stronger over the poles? Are the He\,{\sc i} broad emissions lines
good tracers of the high-velocity wind? Are the Na D and Ca\,{\sc ii} H and K
broad emission lines related to the wind?

Are there preferred stellar latitudes for formation of dust clouds?
If so, one supposes that dust clouds and their accompanying gas
reside in a torus about the star. Evidence for asymmetrical
distributions of dust about RCBs has been discussed previously (see,
for example, Clayton 1996; Rao et al. 1999)
If there is a torus, which broad (presumably) emission lines arise there? 
One may suspect that dust formation may be inhibited at locations 
from which a fast wind originates. Thus, a dust torus and polar winds
may be a natural pairing.

An extended region seems demanded to account for the sharp emission lines.
Is this region restricted to cetain latitudes and locations?

These questions of geometry must be paired with corresponding questions
about the excitation of the lines which lead into further questions
about the heating and cooling of the gas and dust. All of these and the
above questions may be answered by additional observations of RCBs in
decline. This paper certainly helps to suggest which spectroscopic
signatures may be unchanged or little changed and which can be
appreciably changed from one decline to another decline of R CrB.
To define the full range of decline to decline differences (or the
lack thereof) across a sample of RCBs will be a herculean task.

\section{Acknowledgements}
                                                                                  
   We would like to thank Yu. Efimov for his generosity in supplying
  his photometric observations of 2003 minimum before publication.
  We also would like to thank the AAVSO  for permission to use their 
photometric data.
 Some data were also taken from
the website  http://vsnet.kusastro.kyoto-u-ac.jp/vsnet.
  We greatfully appreciate the  help of Gajendra Pandey in producing the
figures.
                                                                                     
The Hobby-Eberly Telescopes (HET) is a joint project of the University of Texas at
Austin, the Pennsylvania State University, Stanford University, 
Ludwig-Maximilians-Universit\"{a}t M\"{u}nchen, and Georg-August-Universit\"{a}t
G\"{o}ttingen. The HET is named in honour of its principal benefactors
W.P. Hobby and Robert E. Eberly.

\end{document}